\documentclass[preprintnumbers,11pt,nofootinbib]{revtex4}

\usepackage{amsmath,latexsym,amssymb,amsfonts}
\usepackage[dvips]{graphicx,color}
\usepackage{bm}

\begin{document}

\preprint{YITP-14-25}

\title{Thin-shell bubbles and information loss problem in anti de Sitter background}

\author{\textsc{Misao Sasaki$^{1,2}$}\footnote{{\tt misao{}@{}yukawa.kyoto-u.ac.jp}} 
and \textsc{Dong-han Yeom$^{1,3}$}\footnote{{\tt innocent.yeom{}@{}gmail.com}}
}

\affiliation{$^{1}$\small{Yukawa Institute for Theoretical Physics, 
Kyoto University, Kyoto 606-8502, Japan}
\\
$^{2}$\small{Tomsk State Pedagogical University, 634050 Tomsk, Russia}\\
$^{3}$\small{Leung Center for Cosmology and Particle Astrophysics, National Taiwan University, Taipei 10617, Taiwan}
}

\begin{abstract}
We study the motion of thin-shell bubbles and their tunneling in anti 
de Sitter (AdS) background. 
We are interested in the case when the outside of a shell
is a Schwarzschild-AdS space (false vacuum) and the inside of it
is an AdS space with a lower vacuum energy (true vacuum). 
If a collapsing true vacuum bubble is created, classically it will form a 
Schwarzschild-AdS black hole. 
However, this collapsing bubble can tunnel to a bouncing bubble that
moves out to spatial infinity. Then, although the classical causal 
structure of a collapsing true vacuum bubble has the singularity and the event
horizon, quantum mechanically the wavefunction has support for a history without
any singularity nor event horizon which is mediated by the non-perturbative,
quantum tunneling effect.
This may be regarded an explicit example that shows the unitarity of an
asymptotic observer in AdS, while a classical observer who only follows 
the most probable history effectively lose information due to the formation
of an event horizon.
\end{abstract}

\maketitle




\section{Introduction}

One of the most interesting issues in black hole physics is the information 
loss problem \cite{Hawking:1976ra}. It is related to the unitarity and the 
fundamental predictability of quantum gravity. 
After a black hole evaporates via Hawking radiation \cite{Hawking:1974sw},
can we restore information from the Hawking radiation or not?
To our best knowledge, the only situation in which the unitarity is believed
to be maintained is the case of asymptotically anti de Sitter (AdS) space
via AdS/CFT correspondence \cite{Maldacena:1997re}, 
though the detailed mechanism has not been well understood.

There are a number of candidates that try to explain the information loss 
problem by a constructive way. First, let us summarize them with comments 
on their potential problems.
\begin{enumerate}
\item \textit{Information loss} exists~\cite{Hawking:1976ra}. 
In this case we encounter a serious problem in the formulation
of quantum field theory \cite{Banks:1983by}.
\item \textit{Black hole complementarity} holds \cite{Susskind:1993if}, and 
information will be recovered by Hawking radiation \cite{Page:1993wv}. 
But, there are some counterexamples \cite{Yeom:2008qw,Hwang:2012nn,Kim:2013fv} 
that show its inconsistency. Namely, the black hole complementarity allows the 
explicit duplication of information. A recent discussion given 
in \cite{Almheiri:2012rt} is in the same line and also shows the inconsistency.
\item Information remains in \textit{small remnants} at the final stage of 
the evaporation. This is connected to 
the \textit{regular black hole picture} \cite{Ashtekar:2005cj}.
However, this would violate the area-entropy relation $\log N = A/4$, where $N$ 
is the number of states and $A$ is the area of the black hole.
That is, this scenario naturally leads the situation that a small remnant,
presumably a Planck size object, carries a huge amount of 
information \cite{Giddings:1993km}. Can this be possible?
\item 
Information is stored in
\textit{large objects} \cite{Giddings:1992hh}, for example 
in \textit{fuzzballs} \cite{Mathur:2005zp} or 
at \textit{firewalls} \cite{Almheiri:2012rt}. 
In this picture it is difficult to imagine what would happen to 
an in-falling observer. Would an in-falling observer be destroyed? If so, 
what would be its effects to an asymptotic observer \cite{Kim:2013fv}? 
If not, then is it free from the problem of the black hole 
complementarity \cite{Hwang:2012nn}?
\item Information disappears to other universes through the
creation of a false vacuum bubble inside the 
event horizon~\cite{Farhi:1989yr,Yeom:2009mn}. 
This is an interesting possibility but does not seem to have
sufficient generality.
\item Information is stored in the form of 
\textit{quantum entanglements} between the outgoing Hawking radiation
and the infalling matter~\cite{Horowitz:2003he}.
This may be possible under some special assumptions.
But up to now, no rigorous idea has been proposed that may cover general 
situations \cite{Gottesman:2003up,Hong:2008ga}.
\item There is \textit{effective loss} of 
information \cite{Maldacena:2001kr,Hawking:2005kf}, 
but not at the fundamental level.
This means that there is no classical observer who can read information 
from Hawking radiation, although information can be recovered by an ideal 
asymptotic observer. Then, what is the difference between the classical 
observer and the asymptotic observer? Can there be an explicit example?
\end{enumerate}

In this paper, we focus on this last possibility.
Namely, the idea is that to understand the information loss problem, 
one has to include not only perturbative effects (e.g., Hawking radiation), 
but also non-perturbative effects that contribute to the entire wavefunction. 
However, because no explicit example has been constructed the physical
meaning of the non-perturbative effects was not clear. 
Can information be encoded in Hawking radiation or in a completely different
form?

To deal with this problem, in this paper, we study 
the motion of a thin-shell vacuum bubble and its quantum tunneling in AdS space. 
We impose the spherical symmetry and assume that both inside and outside
of a shell are described by Schwarzschild-AdS space but with different mass
and vacuum energy. For the interior of a shell, we set the mass parameter
zero so that it is a pure AdS space (true vacuum).
For the exterior of a shell, we assume a positive mass parameter and 
a vacuum energy higher the true vacuum (false vacuum).

In this situation, there are three possible types of classical solutions 
for the motion of a shell: time-symmetric expanding and collapsing solutions
(referred to as symmetric collapsing solutions below for simplicity), 
time-symmetric bouncing solutions (bouncing solutions), 
and time-asymmetric solutions which either expand from the past singularity to 
the boundary or vice versa \cite{Freivogel:2005qh}. 
Which types of solutions are allowed depends on the model parameters.
If a setup allows both symmetric collapsing and bouncing solutions,
then a collapsing bubble can tunnel to a bouncing bubble.

The tunneling of a collapsing vacuum bubble was already investigated 
by Farhi, Guth, and Guven, as well as Fischler, Morgan, and 
Polchinski \cite{Farhi:1989yr}. However, their interest was
in false vacuum bubbles that would induce inflation.
Recently Gregory, Moss, and Withers \cite{Gregory:2013hja} observed that 
tunneling from a collapsing bubble to an expanding bubble can mediate 
the disappearance of a black hole.
However they only considered the de Sitter background and 
hence there was a bound on the mass parameter.
If the mass is bounded, then perturbative effects (Hawking radiation
and its backreaction) may dominate and hence the physical
significance of the tunneling solution becomes unclear. 
In contrast, in the AdS background, one can consider a sufficiently 
large mass parameter which gives an eternal black hole
so that the perturbative effects are always small.
Then the non-perturbative effects become essential.

Thus, we will show that the thin-shell tunneling in AdS has an
 important meaning on the information loss problem. We argue that this 
is an explicit example for the seventh possibility listed above: 
\textit{effective loss of information}. 
We do not claim that it may be applicable to
all possible cases of the black hole evaporation.
Nevertheless, this explicit construction shows the exact meaning 
of the effective loss of information: information is restored for the asymptotic 
unitary observer while the usual classical (most probable) observer 
loses information. Furthermore, the difference between the unitary observer
 and the classical observer resolves the contradiction between the 
unitarity and equivalence principle, which was the motivation 
for the introduction of the firewall \cite{Almheiri:2012rt}.

The paper is organized as follows.
In Section~\ref{sec:thi}, we discuss details of the conditions that allow 
tunneling from a collapsing bubble to a bouncing bubble. 
In Section~\ref{sec:tun}, we consider the tunneling and its causal structure,
as well as the physical meaning in the light of the information loss problem. 
Finally, in Section~\ref{sec:con}, we summarize our results and discuss 
possible future directions.

\section{Thin-shell bubbles in AdS space}
\label{sec:thi}

In this section, we construct thin-shell bubble solutions in the AdS background.

\subsection{Junction equations}

We consider spacetime with the metric,
\begin{eqnarray}
\label{eq:metric}
ds_{\pm}^{2}= - f_{\pm}(R) dT^{2} + \frac{1}{f(R)} dR^{2} + R^{2} d\Omega^{2}\,;
\quad f_{\pm}(R)= 1 - \frac{2M_{\pm}}{R} + \frac{R^{2}}{\ell_{\pm}^{2}}\,,
\end{eqnarray}
where the suffices $\pm$ denote the exterior $(+)$ and interior $(-)$ of 
a thin-shell. 
The AdS radius $\ell_\pm$ is related to the vacuum energy density
$\rho_\pm (<0)$ as
\begin{eqnarray}
\ell^{2}_{\pm} = \frac{3}{8\pi |\rho_{\pm}|}\,.
\end{eqnarray}
If $\ell_{-} < \ell_{+}$, it is a true vacuum bubble,
and if $\ell_{-} > \ell_{+}$, it is a false vacuum bubble. 
In this paper, we are interested in the case of a true vacuum bubble,
so we assume $\ell_{-} < \ell_{+}$.
In addition, we consider the case when the inside is a pure AdS space
(pure true vacuum). Hence we assume $M_{+} >0$ and set $M_{-}=0$.

We denote the radius of the shell by $r$.
Then one can express the intrinsic metric on the thin-shell as
\begin{eqnarray}
ds^{2} = - dt^{2} + r^{2}(t) d\Omega^{2}\,.
\end{eqnarray}
The equation of motion of the thin-shell is determined by the 
junction condition \cite{Israel:1966rt}:
\begin{eqnarray}\label{eq:junc}
\epsilon_{-} \sqrt{\dot{r}^{2}+f_{-}(r)} 
- \epsilon_{+} \sqrt{\dot{r}^{2}+f_{+}(r)} = 4\pi r \sigma\,,
\end{eqnarray}
where $\sigma$ is the surface tension which is assumed to be positive,
and $\epsilon_\pm$ are the signs of the extrinsic curvature of the shell 
in the two-dimensional $(t,r)$-space.
Namely, $\epsilon_\pm = + 1$ if $R$ increases along the outward normal 
of the shell and $\epsilon_\pm = - 1$ if $R$ decreases along the outward normal.

\subsection{Signs of extrinsic curvature and causal structure}

\begin{figure}
\begin{center}
\includegraphics[scale=0.7]{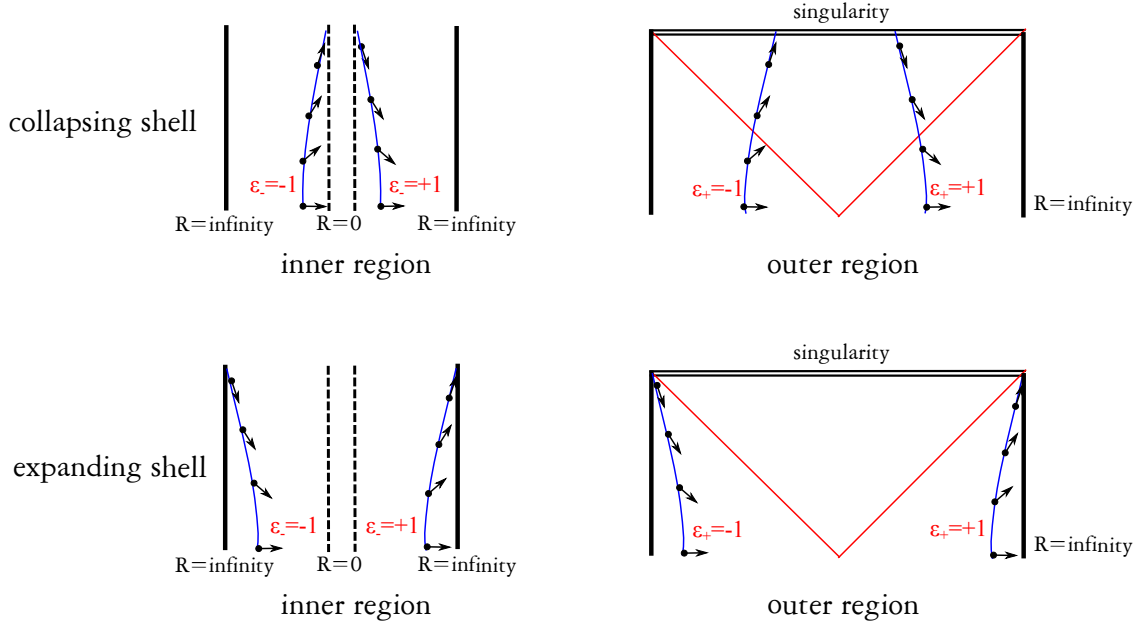}
\caption{\label{fig:signs}
Signs of $\epsilon_{\pm}$ for collapsing and expanding shells.}
\end{center}
\end{figure}

Figure~\ref{fig:signs} summarizes the relation between the signs of 
$\epsilon_{\pm}$ and the causal structure. 
The upper figures describe the collapsing shell and the lower figures 
the bouncing shell. The blue curves are the possible thin-shell 
trajectories, and we adopt the convention that the right-hand side
of the shell is the outside. Therefore, a thin-shell bubble spacetime
is constructed by matching the right-hand side of the blue curve in the 
right figure with the left-hand side of the blue curve in the left figure.
The black arrows are the direction of the outward normal direction.

A short manipulation of the junction condition gives
a simpler formula:
\begin{eqnarray}
\dot{r}^{2} + V(r) &=& 0\,;
\label{eq:form}
\\
V(r) &=& f_{+}(r)- \frac{\left(f_{-}(r)-f_{+}(r)
-16\pi^{2} \sigma^{2} r^{2}\right)^{2}}{64 \pi^{2} \sigma^{2} r^{2}}\,
\cr\cr
&=&
1 - \frac{M_{+}^{2}}{16 \pi^{2} \sigma^{2} r^{4}} 
- \frac{\left( 32\pi^{2} \sigma^{2}
+ \mathcal{A} \right)M_{+}}{16 \pi^{2} \sigma^{2} r} 
- \left( \mathcal{A}^{2}- \frac{64 \pi^{2} \sigma^{2}}{\ell_{+}^{2}} \right)
 \frac{r^{2}}{64 \pi^{2} \sigma^{2}},
\label{eq:Veff}
\end{eqnarray}
where
\begin{eqnarray}
\mathcal{A} 
\equiv \frac{1}{\ell_{-}^{2}} - \frac{1}{\ell_{+}^{2}} - 16 \pi^{2} \sigma^{2}.
\end{eqnarray}
Thus the problem reduces to a simple one-dimensional motion with 
the effective potential $V(r)$. However, the price to pay
is the loss of information about the signs of $\epsilon_{\pm}$. 

To recover the signs, we have to check the signs of the extrinsic curvatures:
\begin{eqnarray}\label{eq:ec1}
\beta_{+}(r) &\equiv&
 \frac{f_{-}(r)-f_{+}(r)-16\pi^{2} \sigma^{2} r^{2}}{8 \pi \sigma r}
 = \epsilon_{+} \sqrt{\dot{r}^{2}+f_{+}(r)}\,,
\\ \label{eq:ec2}
\beta_{-}(r) &\equiv&
 \frac{f_{-}(r)-f_{+}(r)+16\pi^{2} \sigma^{2} r^{2}}{8 \pi \sigma r}
 = \epsilon_{-} \sqrt{\dot{r}^{2}+f_{-}(r)}\,.
\end{eqnarray}
Thus the signs of $\beta_\pm$ determine those of $\epsilon_\pm$.
Since we assume $\sigma>0$, the possible cases are:
\begin{list}{}{}
\item{(1)} $\epsilon_+=+1$ and $\epsilon_-=+1$.
\item{(2)} $\epsilon_+=-1$ and $\epsilon_-=+1$.
\item{(3)} $\epsilon_+=-1$ and $\epsilon_-=-1$.
\end{list}
An immediate consequence of our assumption that $\sigma>0$
is $\beta_{\pm}\to M_+/(4\pi\sigma r^2)>0$ as $r\to0$.
Hence $\epsilon_{\pm}$ must be positive in the limit $r\to0$. 
In addition, since $\ell_{-}<\ell_{+}$, we see $\beta_{-}>0$ 
as $r\to\infty$. Thus $\epsilon_{-}=+1$ for both $r\to0$ and $r\to\infty$.
This excludes the cases (3) for both symmetric collapsing and bouncing solutions.
As we will discuss below, since we are interested in the case when
the quantum tunneling may prevent the appearance of an event horizon, 
the case (2) is also excluded. This leaves only the case (1) as relevant.

\subsection{Existence of solution}

Because of Eq.~(\ref{eq:form}), the only allowed region is where $V(r) < 0$. 
Keeping this in mind, we define the symmetric and asymmetric solutions as follows:
\begin{description}
\item[Symmetric solution:] If there is a point $r_*$ at which
$V(r_*)=0$ and either $V(r < r_*) < 0$ or $V(r > r_*) < 0$, then 
the allowed regions are either $0 <r \leq r_*$ or $r_* \leq r < \infty$. 
For the former, the shell radius expands from zero, becomes maximum
at $r = r_*$, and collapses to zero again.
For the latter, the shell radius contracts from infinity,
becomes minimum at $r = r_*$, and expands to infinity again. 
Both of these solutions are symmetric under time reversal.
So we call them \textit{symmetric solutions}.
\item[Asymmetric solution:] If there exists no zero point of $V(r)$, we must have
$V(r) < 0$ for all $r$. In this case, either the shell radius increases from 
zero to infinity or decreases from infinity to zero. 
We call them \textit{asymmetric solutions}.
\end{description}

What we are interested in is tunneling from a collapsing bubble to an 
expanding bubble without forming an event horizon. This kind of tunneling 
may happen between symmetric collapsing and symmetric bouncing solutions. 
So we require the existence of both symmetric collapsing and bouncing solutions.
Moreover, we require that the exterior of the expanding shell after tunneling
should contain the spatial infinity of the Schwarzschild-AdS space.
This means $\epsilon_{+}>0$ in the limit $r\to\infty$.
These requirements are realized by the following conditions:
\begin{enumerate}
\item $V(r)$ has a unique maximum at $r=r_0$ at which $V(r_0)>0$. 
\item For $0<r<r_0$, $V(r)$ is monotonically increasing,
 and $V(r)<0$ as $r\to0$.
\item For $r_0<r<\infty$, $V(r)$ is monotonically decreasing,
 and $V(r)<0$ as $r\to\infty$.
\item $\beta_{+}(r)>0$ as $r\to\infty$.
\end{enumerate}
The first three conditions imply the existence of two zero points, say 
$r^{(a)}$ and $r^{(b)}$ (where $r^{(a)}<r^{(b)}$) such that $V'(r^{(a)})>0$
and $V'(r^{(b)})<0$. Below we examine these conditions.

\subsubsection{$r \rightarrow 0$ limit}

For $r \rightarrow 0$ limit, $V(r)$ is expanded by
\begin{eqnarray}
V(r\rightarrow 0) \simeq - \frac{M_{+}^{2}}{16 \pi^{2} \sigma^{2} r^{4}} 
+ \mathcal{O}\left(r^{-1}\right).
\end{eqnarray}
Therefore $V(r\rightarrow 0)<0$ is always guaranteed for $M_{+} \neq 0$ $(>0)$.

\subsubsection{$r \rightarrow \infty$ limit}

For $r \rightarrow \infty$ limit, $V(r)$ is expanded by
\begin{eqnarray}
V(r\rightarrow \infty) \simeq 
\left(\frac{1}{\ell_{+}^{2}} - \frac{\mathcal{A}^{2}}{64 \pi^{2} \sigma^{2}}\right)
r^{2} + 1 + \mathcal{O}\left(r^{-1}\right).
\end{eqnarray}
Therefore, the condition $V(r\rightarrow \infty) < 0$ means 
\begin{eqnarray}\label{eq:cond1}
 64 \pi^{2} \sigma^{2}\ell_{+}^2<\mathcal{A}^{2} \ell_{+}^4
=\left(\frac{\ell_{+}^2}{\ell_{-}^2}-1-16\pi^2\sigma^2\ell_{+}^2\right)^2\,.
\end{eqnarray}
In other words,
\begin{eqnarray}\label{eq:cond2}
\ell_{-} <
 \frac{\ell_+}{\sqrt{1+8 \pi \sigma\ell_{+}+16 \pi^{2} \sigma^{2}\ell_+^2}}
\equiv \ell_{\mathrm{c}}
\quad\mbox{for}~\mathcal{A}>0\,,
\end{eqnarray}
or
\begin{eqnarray}\label{eq:cond2x}
\frac{\ell_+}{\sqrt{1-8 \pi \sigma\ell_{+}+16 \pi^{2} \sigma^{2}\ell_+^2}}
<\ell_{-}\quad\mbox{for}~~\mathcal{A}<0\,.
\end{eqnarray}

Here we consider the condition that $\beta_{+}>0$ as $r\to\infty$.
We have
\begin{eqnarray}
\beta_{+}(r)\mathop{\to}\limits_{r\to\infty}\frac{r}{8\pi\sigma}\mathcal{A}+\mathcal{O}(1)\,.
\end{eqnarray}
Thus we must require $\mathcal{A}>0$.
Hence Eq.~(\ref{eq:cond2}) is the relevant condition.
Here we note that $\mathcal{A}>0$ is equivalent to
\begin{eqnarray}
\label{eq:cond3}
\ell_{-} < \frac{\ell_{+}}{\sqrt{1+16 \pi^{2} \sigma^{2}\ell_{+}^2}}\,.
\end{eqnarray}
Comparing this with Eq.~(\ref{eq:cond2}), we see that the positivity
of $\mathcal{A}$ is automatically guaranteed if $\ell_{-}<\ell_{\mathrm{c}}$.

\begin{figure}
\begin{center}
\includegraphics[scale=0.75]{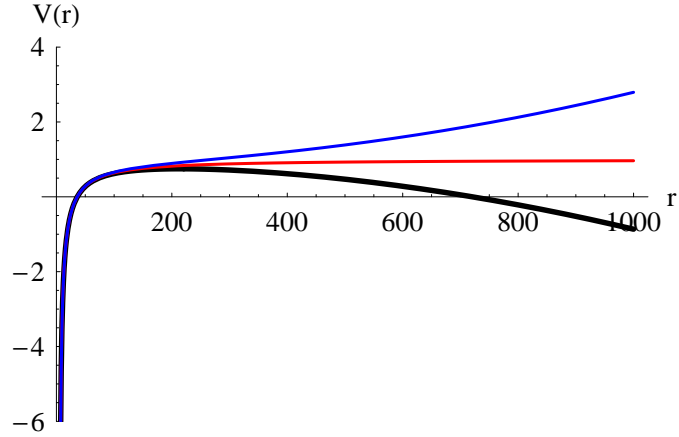}
\caption{\label{fig:VAdS}The effective potential $V(r)$ in the vacuum bubbles
in AdS background, for $M_{+} = 10$, $\ell_{+}=10$, and $\sigma=0.01$ in
Planck units.
If $\ell_{-}$ saturates the condition~(\ref{eq:cond2}), i.e., 
$\ell_{-} = \ell_{\mathrm{c}}$ (red curve), $V(r)$ asymptotically 
approaches unity. 
If $\ell_-$ is slightly greater than $\ell_{\mathrm{c}}$,
say $\ell_-=\ell_{\mathrm{c}}+10^{-4}$ (blue curve),
or smaller than $\ell_{\mathrm{c}}$, say $\ell_-=\ell_{\mathrm{c}}-10^{-4}$
 (black curve), the large $r$ behavior changes.}
\end{center}
\end{figure}

\subsubsection{Around the maximum}

If Eq.~(\ref{eq:cond1}) is satisfied and $M_{+}>0$, then there exists
a point $r_{0}$ at which $V'(r_{0})=0$. Taking the derivative of $V(r)$ given
in Eq.~(\ref{eq:Veff}), it turns out that $V'(r)=0$ gives
a quadratic equation for $r^3$ with either positive and negative roots.
Since $r$ must be positive, this means the point $r_0$ is unique if it exists.
Hence $V(r)$ is monotonically increasing for $0<r<r_0$ and monotonically
decreasing for $r_0<r<\infty$. The value at the maximum $V(r_{0})$ 
may be expressed as
\begin{eqnarray}
V(r_{0}) = f_{+} + \frac{r_{0}}{2}f'_{+} 
- \frac{\left( \mathcal{A}r_{0}^{3} + 2M_{+}\right)
\left( \mathcal{A}r_{0}^{3} - M_{+}\right)}{32 \pi^{2} \sigma^{2} r_{0}^{4}}\,,
\end{eqnarray}
where $f'_{+} = 2M_{+}/r_{0}^{2} + 2 r_{0}^{2}/\ell_{+}^{2}$ and 
$f'_{-} = 2 r_{0}^{2}/\ell_{-}^{2}$.

Note that for the critical case $\ell_{-}=\ell_{\mathrm{c}}$
($\mathcal{A} = \mathcal{A}_{\mathrm{c}}\equiv 8 \pi \sigma/\ell_{+}$), 
$V(r)\to+1$ as $r\to\infty$ (see Fig.~\ref{fig:VAdS}). 
Therefore, if the value of $\ell_{-}$ (or $\mathcal{A}$)
is very close to the critical limit, 
the positivity of the maximum $V(r_{0})$ is guaranteed.

\begin{figure}
\begin{center}
\includegraphics[scale=0.5]{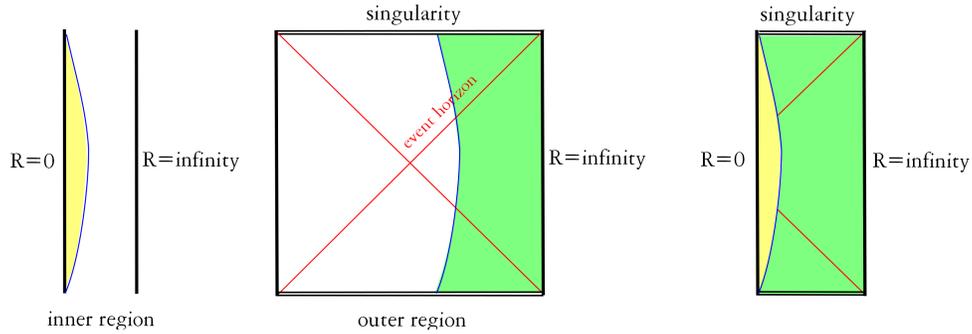}
\caption{\label{fig:AdSbubble}Causal structure of a collapsing AdS bubble.}
\end{center}
\end{figure}
\begin{figure}
\begin{center}
\includegraphics[scale=0.5]{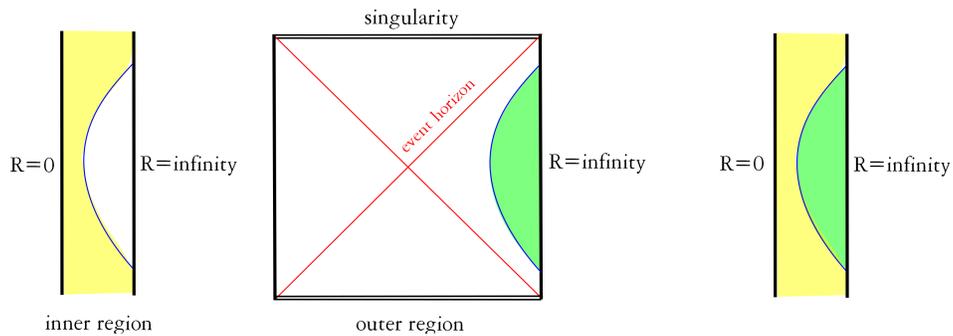}
\caption{\label{fig:AdSbubble2}Causal structure of a bouncing AdS bubble.}
\end{center}
\end{figure}
\begin{figure}
\begin{center}
\includegraphics[scale=0.5]{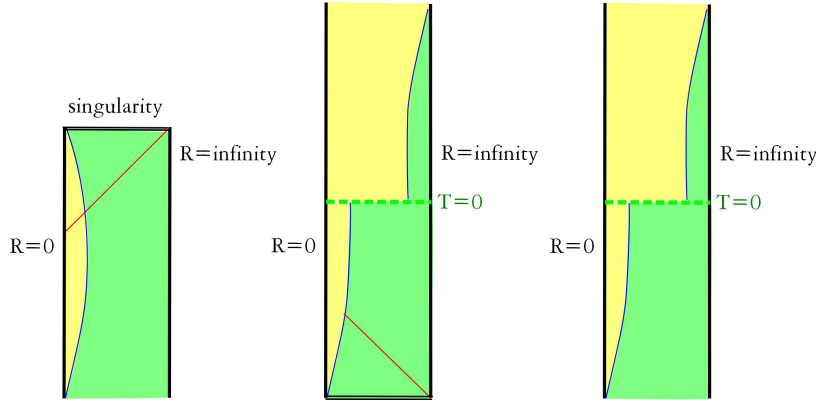}
\caption{\label{fig:AdStunnel}Left: Usual classical evolution. 
Center: Our model for tunneling from a collapsing bubble to an expanding 
bubble. Right: A realistic case of tunneling from a collapsing bubble to 
an expanding bubble.}
\end{center}
\end{figure}

This condition can be intuitively understood as follows.
If the effect of gravity is relatively weak, 
the energy of the system can be approximated by three factors,
the kinetic energy due to the velocity of the shell, the contribution of the 
tension, and the vacuum energy difference.
Since we are considering the tunneling process, we may focus on
turning points where the kinetic energy is zero. For the existence of
the tunneling, there should be two turning points.
The radius of a turning point
should approximately satisfy the following relation:
\begin{eqnarray}
4\pi r^{2} \sigma - \frac{4\pi}{3} 
\left( \left| \rho_{-} \right| - \left| \rho_{+} \right| \right) r^{3} \simeq M_{+},
\end{eqnarray}
or equivalently,
\begin{eqnarray}
r^{3}- \frac{8\pi \sigma}{\ell_{-}^{-2} - \ell_{+}^{-2}}r^2
- \frac{2M_{+}}{\ell_{-}^{-2} - \ell_{+}^{-2}}\approx0\,.
\end{eqnarray}
This implies that for the true vacuum case ($\ell_{-}^{-2} - \ell_{+}^{-2} > 0$), 
it is possible to have two solutions for $r>0$. When the tension is small enough,
this corresponds the condition that $\ell_{-} < \ell_{+} \sim \ell_{\mathrm{c}}$. 
Hence, our condition $\ell_{-} < \ell_{\mathrm{c}}$ physically means that,
when the bubble tunnels to a larger one, the inner true vacuum should be 
\textit{deep enough} to compensate the increase in the energy contribution 
from the tension.

\vspace{10mm}

To summarize, for any given $M_{+}$, $\ell_{+}$, and $\sigma$,
if we choose $\ell_{-}$ to satisfy the condition~(\ref{eq:cond2}) with the
value almost saturating it, that is, $\ell_{-} \lesssim \ell_{\mathrm{c}}$, 
the value of $\mathcal{A}$ becomes close enough to its critical value so that
$V(r_0)>0$ is guaranteed, one can always find these two kinds of
(i.e. symmetric collapsing and bouncing) solutions.
The causal structure of a symmetric collapsing shell is depicted in 
Fig.~\ref{fig:AdSbubble} and that of a bouncing shell in 
Fig.~\ref{fig:AdSbubble2}. In these figures, the yellow colored regions
correspond to the interior of the shell and the green colored
regions are the exterior of the shell. In both figures the one on the right
describes the bubble shell spacetime constructed by pasting
the yellow region on the left and the green region in the middle together.

\section{Tunneling to a bouncing shell}
\label{sec:tun}

Let us assume that the model parameters are such that they allow the
existence of both collapsing and bouncing solutions as discussed in the 
previous section. We now discuss the quantum tunneling between these 
two solutions.

\subsection{Tunneling instanton}

Let us first recall the causal structures of the symmetric collapsing 
and bouncing solutions as shown in the rightmost ones in 
Figs.~\ref{fig:AdSbubble} and~\ref{fig:AdSbubble2}, respectively. 
A tunneling solution would join these two spacetimes on the
surface of time-symmetry, $t=0$. In the rigorous sense, the resulting spacetime 
describes tunneling from a shell expanding from a white hole 
to a bouncing shell expanding to infinity.
Nevertheless, as done in the original paper by Hawking~\cite{Hawking:1974sw},
we may regard the expanding shell with the white hole structure (i.e.
with a past horizon) as an analytical approximation to a more complicated, 
possibly non-analytical situation in which the spacetime is initially
perfectly regular without any past horizon.
Assuming the validity of this picture, Fig.~\ref{fig:AdStunnel}
shows the causal structure of a realistic 
(so-called `buildable' \cite{Freivogel:2005qh}) spacetime endowed
with the tunneling to a larger bouncing bubble.

\begin{figure}
\begin{center}
\includegraphics[scale=0.75]{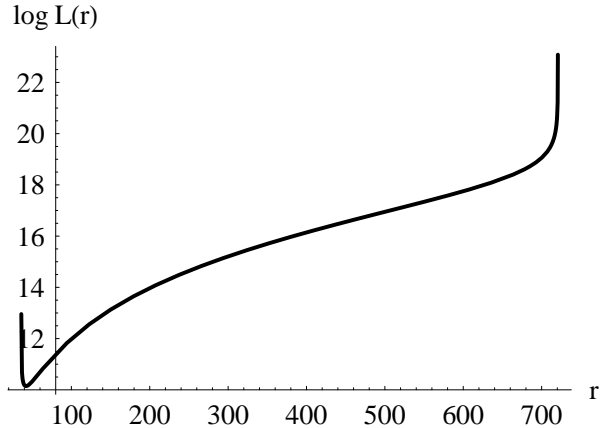}
\caption{\label{fig:Lftn}$\log L(r)$ for the example of 
$M_{+} = 10$, $\ell_{+}=10$, $\sigma=0.01$, and 
$\ell_{-}=\ell_{\mathrm{c}}-10^{-4}$. 
$L(r)$ diverges at the two turning points $r^{(a)}\simeq 36.14$ 
and $r^{(b)}\simeq 720.74$, though the integration converges.}
\end{center}
\end{figure}

At leading order, we may regards the tunneling to be mediated by
instanton solutions, that is, solution in Euclideanized time $\tau=it$.
Let $r=r(\tau)$ be the instanton that dominates the tunneling.
Then the tunneling probability may be evaluated as $\Gamma \sim \exp[-B]$, 
where $B$ is the instanton action given by~\cite{Ansoldi:1997hz}
\begin{eqnarray}
B = \int_{r^{(a)}}^{r^{(b)}} L(r) dr\,,
\end{eqnarray}
where
\begin{eqnarray}
L(r) = r^{2} \left[ \frac{64 \pi^{2} \sigma^{2} r^{2} f_{-}f_{+} 
+ \left( f'_{-} +f'_{+} \right)f_{-}f_{+} 
- \left(f_{-}^{2} f'_{+} + f'_{-} f_{+}^{2} - 16\pi^{2}\sigma^{2}r^{2} 
\left(f'_{-}f_{+}+f_{-}f'_{+} \right)\right)}{2f_{-}f_{+}\sqrt{4f_{-}f_{+}
-\left(f_{-}+f_{+}-16\pi^{2} \sigma^{2} r^{2}\right)^{2}}} \right].
\end{eqnarray}
This is plotted in Fig.~\ref{fig:Lftn}. As seen from it
$L(r)$ diverges at the turning points, $V(r^{(a)}) =V(r^{(b)})= 0$. 
However, the integration is finite, since it is in the form,
\begin{eqnarray}
B=\int_{r^{(a)}}^{r^{(b)}} \frac{(\mathrm{finite\;\; terms})}{\sqrt{V(r)}}dr
 = \int_{\tau^{(a)}}^{\tau^{(b)}} 
\frac{(\mathrm{finite\;\; terms})}{\sqrt{V(r)}} \frac{dr}{d\tau}d\tau
=\int_{\tau^{(a)}}^{\tau^{(b)}}(\mathrm{finite\;\; terms})d\tau\,,
\end{eqnarray}
where the last equality follows from the Euclidean equation of motion,
$dr/d\tau =\pm \sqrt{V(r)}$.

\subsection{Interpretation}

The most probable history is the spacetime that describes 
usual gravitational collapse and formation of a black hole,
as depicted on the left in Fig.~\ref{fig:AdStunnel}.
In this case, the black hole may become stable if 
$\ell_+\lesssim M_+$~\cite{Hawking:1982dh}. This means even if we include 
the semi-classical effect of Hawking radiation, the causal structure 
does not change. Namely, the black hole becomes eternal. 
Then after an infinite lapse of time, all correlations between inside and
outside the event horizon will eventually be destroyed~\cite{Maldacena:2001kr}.
However, since the lifetime is infinite, even if the non-perturbative 
tunneling effect is exponentially suppressed, it cannot be neglected if
it exists. Since there is no future event horizon nor singularity, 
correlations among any points will remain forever.
The right figure in Fig.~\ref{fig:AdStunnel} shows such an example. 

To make our argument more relevant in the context of AdS/CFT correspondence, 
let us consider a unitary observer at infinity, which we call 
the unitary boundary observer.
When this observer computes CFT correlations at boundary, he/she
gathers all possible contributions in the path integral. 
Among them, we have at least two solutions: one is the 
classical AdS black hole at thermal equilibrium 
(left of Fig.~\ref{fig:AdStunnel}) with probability $p_{1} \lesssim 1$, 
while the other is the quantum mechanical bouncing AdS bubble 
(right of Fig.~\ref{fig:AdStunnel}) with probability 
$p_{2} \sim e^{-B} \ll 1$. 
For a classical boundary observer who sees the most probable history, 
the black hole exists forever and information will be swallowed
by the black hole. Therefore, it becomes impossible for this observer
to gather information from Hawking radiation, and hence he/she will
see a thermal state in the end.
 On the other hand, a boundary observer for the spacetime
which underwent tunneling can compute CFT correlations without any
difficulty, since the entire causal structure in the bulk is trivial.
In terms of a two-point function, what we have discussed in the above
may be expressed as
\begin{eqnarray}
\langle \phi \phi \rangle \simeq 
p_{1} \langle \phi \phi \rangle_{1} + p_{2} \langle \phi \phi \rangle_{2}
 + \cdots \mathop{\simeq}\limits_{t\to\infty} \mathrm{const} \times e^{-B}\,.
\end{eqnarray}

In other words, information will be stored in the entire wavefunction, 
but the classical observer who experiences only the most probable history 
will effectively lose information. This leads to a very interesting,
important consequence. The expectation value of the geometry for the 
unitary observer will be given approximately by
\begin{eqnarray}
\langle g_{\mu\nu} \rangle 
\simeq p_{1} g_{\mu\nu}^{(1)} + p_{2} g_{\mu\nu}^{(2)} +\cdots\,.
\end{eqnarray}
Apparently the geometry for the unitary observer $\langle g_{\mu\nu} \rangle$ 
is different from that of the classical observer $g_{\mu\nu}^{(1)}$. 
Namely, $\langle g_{\mu\nu} \rangle$ does not satisfy the classical 
Einstein equations while $g_{\mu\nu}^{(1)}$ does. 
This explains why the assumption of the unitarity seems to contradict 
with the laws of general relativity \cite{Almheiri:2012rt}. 
In particular, this suggests
that it is not necessary to introduce the firewall at all.

In fact, the inconsistency with classical general relativity 
and the unitary observer would see has been discussed extensively
in the context of quantum cosmology. In quantum cosmology
a classical geometry would appear only effectively in the 
superspace of 3-geometries where the wavefunction is peaked
along a series of 3-geometries~\cite{Hartle:2007gi}.
Commonly a single wavefunction can contain many classical geometries,
and each classical geometry with separate history is realized 
with different probabilities~\cite{Hartle:2007gi,Hwang:2011mp}. 
Perhaps, the omnipotent (unitary) observer can see every history.
But we, the classical observers, can experience only one of them. 
Like this, in the black hole background, although we may think of
an omnipotent observer who can gather all the information, 
it is not strange that the classical observers lose it.

Of course, our discussion is based on a very simple, toy model. 
So it is premature to say that we now know the detail process of how
the initial information is restored.
In particular, it is impossible to study the evolution of all possible
initial states and the process of information restoration in our model.
To do so, we would need to specify a field theoretical model in detail,
which is beyond the scope of this paper.

However, the information restoration process can be 
discussed by using correlation functions~\cite{Maldacena:2001kr},
at least partially. 
In terms of correlation functions, it is conceptually clear to see the 
restoration of correlations in spacetime with trivial topology. In the end, 
the restored correlation will be exponentially suppressed, which can be 
interpreted as a result of complicated entanglements of the information.
This was discussed by Hawking~\cite{Hawking:2005kf}.

\section{Conclusion}
\label{sec:con}

In this paper, we studied the motion of thin-shell true vacuum bubbles
in anti de Sitter background, and discussed the information loss problem.
 We focused on the case where the exterior of the shell is a 
Schwarzschild-anti de Sitter space and the interior is a pure 
anti de Sitter space, considered quantum tunneling from a collapsing 
bubble to a bouncing bubble.

Since such tunneling is a non-perturbative effect
with a highly suppressed probability, for a classical observer who
experiences the most probable history will see the formation
of a black hole. Hence the boundary observer would lose the information
and find himself/herself in a thermal state. We note that in this case
the spacetime satisfies the classical Einstein equations.

On the other hand, if such tunneling occurs, there will be no event horizon 
nor singularity in the spacetime. 
In this case, the spacetime does not satisfy the classical Einstein
equations, but there is no loss of information because of the trivial topology.
Consequently, an asymptotic observer on the boundary can gather all 
the information. 

Thus we conclude that \textit{information is conserved if the
non-perturbative tunneling effect is taken into account,
while it is effectively lost for a classical observer who follows 
the most probable (i.e. classical) history}. 
This explains why the Einstein equations should be violated for 
the asymptotic \textit{unitary} observer, since he/she should sum-over
all possible histories. In particular, it is unnecessary to introduce
the firewall.

Then the question is whether this kind of tunneling is universal or not.
In this paper, we only considered for symmetric bubbles to
find a well-defined tunneling process.
One other interesting case is an asymmetric collapsing or bouncing bubble.
Then we cannot apply the Euclidean technique to calculate a tunneling probability.
However, quantum mechanically this is a scattering problem so that
there can be a non-zero scattering probability from an incoming wave 
to a bouncing wave,
though the dominant contribution will just reach the singularity and
this will be interpreted as a classical trajectory of a collapsing asymmetric bubble.
Although it is beyond the scope of our paper, if it is possible,
then this will be strengthened our assertions.

On the other hand, if there is no solution that reaches the boundary
(hence, only has symmetric collapsing solution), then it can be difficult
 to impose the tunneling argument. One comment regarding this is that this 
surely requires restricted parameters.
However if the picture of string theory landscape \cite{Susskind:2003kw} 
is correct, one expects almost certainly the existence of a region in the 
landscape where the parameters of theory allow such quantum tunneling. 
Then even if it is a tiny region in the landscape with an exponentially
small probability, the existence of a single example of such tunneling 
is sufficient to recover information.
However, in any case, it is fair to say that we did not prove
the generality of our assertions; rather, the important meaning of our paper 
is that we have shown the effective loss of information, by using 
a \textit{definite example}.

To strengthen our conclusion, there are a few points in our discussion
which need to be improved. One is the thin-shell approximation. 
To make our model more realistic it is desirable if we can introduce
a scalar field and its potential that allows tunneling,
and redo the whole analysis in this Einstein-scalar theory.
Another is the approximation to use the white hole structure as 
the initial state. In reality, we want a perfectly regular initial state
with some concentration of the scalar field energy in the bulk, which
would form a true vacuum bubble and which would collapse to a black hole
classically. 

Related to the above, it seems almost certain that a black hole can form
not only in the sea of false vacuum but also in the true vacuum.
However, to study such a case, we would need to specify a detailed field 
theoretical model, which is again beyond the scope of this paper. 
Here we just would like to emphasize that we have presented a good explicit 
model that demonstrates how the effective loss of information occurs and how
the information can be restored. 

If we can construct realistic initial data in a field theoretical model
and find an instanton that mediates tunneling to an expanding bubble, 
probably we may finally say that the information loss problem has 
been solved, at least in the AdS/CFT context.

\section*{Acknowledgment}

This work is supported by the JSPS Grant-in-Aid for Scientific Research (A) 
No.~21244033.
DY was supported by Leung Center for Cosmology and Particle Astrophysics (LeCosPA) of
National Taiwan University (103R4000).
The authors would like to thank participants of the 
workshop ``APC-YITP collaboration: Mini-Workshop on Gravitation and
 Cosmology'' (YITP-X-13-03) for useful discussion and comments.

\end{document}